\documentclass[aps,prl,reprint,superscriptaddress]{revtex4-1}

\usepackage{url}
\usepackage{color}
\usepackage{graphicx}
\usepackage{bm}
\usepackage{amsmath}
\usepackage{amssymb}
\usepackage{physics}
\usepackage{verbatim}

\begin{document}


\title{Threading the spindle: a geometric study of chiral liquid crystal polymer microparticles}


\author{Helen S. Ansell$^*$}
\affiliation{Department of Physics and Astronomy, University of Pennsylvania, Philadelphia, Pennsylvania 19104, USA}
\author{\firstname{Dae Seok} Kim$^*$}
\affiliation{Department of Physics and Astronomy, University of Pennsylvania, Philadelphia, Pennsylvania 19104, USA}
\affiliation{UMR CNRS 7083, ESPCI PARIS, PSL RESEARCH UNIVERSITY, 75005 PARIS, FRANCE}
\author{Randall D. Kamien}
\author{Eleni Katifori}
\affiliation{Department of Physics and Astronomy, University of Pennsylvania, Philadelphia, Pennsylvania 19104, USA}
\author{Teresa \surname{Lopez-Leon}}
\affiliation{UMR CNRS 7083, ESPCI PARIS, PSL RESEARCH UNIVERSITY, 75005 PARIS, FRANCE}



\date{\today}

\begin{abstract}
Polymeric particles are strong candidates for designing artificial materials capable of emulating the complex twisting-based functionality observed in biological systems. In this letter, we provide the first detailed investigation of the swelling behavior of bipolar polymer liquid crystalline microparticles. Deswelling from the spherical bipolar configuration causes the microparticle to contract anisotropically and twist in the process, resulting in a twisted spindle shaped structure. We propose a model to describe the observed spiral patterns and twisting behavior. 
\end{abstract}

\pacs{}

\maketitle



From the twisting of DNA and proteins to the twirling of plant tendrils and wringing of a wet towel, torsional and twisting motion is an inevitable and common occurrence at all  scales. 
Twisting motion allows many organisms to achieve their complex mechanical functions including swimming~\cite{Arroyo2012}, flying~\cite{Young2009}, crawling~\cite{Marvi2014}, climbing \cite{Isnard2009}, seedpod opening~\cite{Forterre2011}, and energy storage \cite{Mahadevan2000}, so biology has perfected a broad range of ingenious mechanisms to achieve it. The three-dimensional twisting and torsional actuation of biological systems often originates from anisotropic internal microstructures with local volume variation. For example, cucumber tendrils coil and wind via asymmetric contraction of an internal fiber ribbon of specialized cells~\cite{Gerbode2012,Wang2013} and euglenas swim by uniformed sliding and twisting of their cell walls 
~\cite{Noselli2019}. Emulation of these types of highly controlled twisted shape transformations in nature would provide the capacity and incentive to design active materials with specific twisting-based functionality.
 
Polymeric particles are one of the strongest candidates for usage in a diverse range of applications including drug delivery~\cite{Pekarek1994}, emulsion stabilization~\cite{Madivala2009}, catalysis~\cite{Cui2013}, separation processes~\cite{Davis2002}, and sensors~\cite{Gider2007}. The shape complexity, size, and internal morphology of the particles are critical for their specific applications. Liquid crystal polymers (LCPs) can be considered as particularly favorable due to the possibility of designing the structure of the polymer network by templating using the nematic director field. Changes in the shape and structure of the particles, including the emergence of chirality from an initially achiral state, can be driven by changes in external stimuli including heat~\cite{Ware2015}, pH~\cite{Klinger2014}, light~\cite{Oosten2009}, electric field~\cite{Jager2000}, magnetic field~\cite{Tang2015}, and solvents~\cite{Kamal2014a}. 

LCP microparticles formed from bipolar nematic droplets provide a beautiful example of how an initially achiral configuration can lead to a chiral final configuration. Volume contraction, driven by the removal of non-reactive mesogens from the polymeric particles, causes the spherical microparticle to contract anisotropically and twist in the process, resulting in a twisted spindle shaped structure~\cite{Lee2014,Wang2016}. The mechanism through which the twisting occurs as well as the role of the bipolar configuration and the interplay between geometric frustration and bulk instabilities in initially curved droplets remain poorly understood. In this letter, we report the first detailed investigation into the behavior of these twisted particles. By carefully controlling the swelling of the microparticles using a binary solvent mixture~\cite{Kamal2014a}, we are able to stabilize the intermediate states between the initial, achiral spherical structure and the final, twisted, spindle structure, and measure the relationship between the aspect ratio of the microparticle \(u\) and twist angle of the polymer strands at the surface \(\beta\). We find that the microparticles deswell following a two-step process: below a critical  \(u\) value, they shrink predominantly by inner folding of the polymer bundles, while above the critical value, they shrink mostly by twisting. We show that the particles twist from the surface inwards by aligning their polymer strands along loxodromes, or lines of constant bearing. We propose a model to describe the observed spiral patterns and twisting behavior. Our results suggests a strategy to design chiral spindle-shaped particles with tunable twist and sensing or actuation capabilities.

\begin{figure}
		\includegraphics[width=0.49\textwidth]{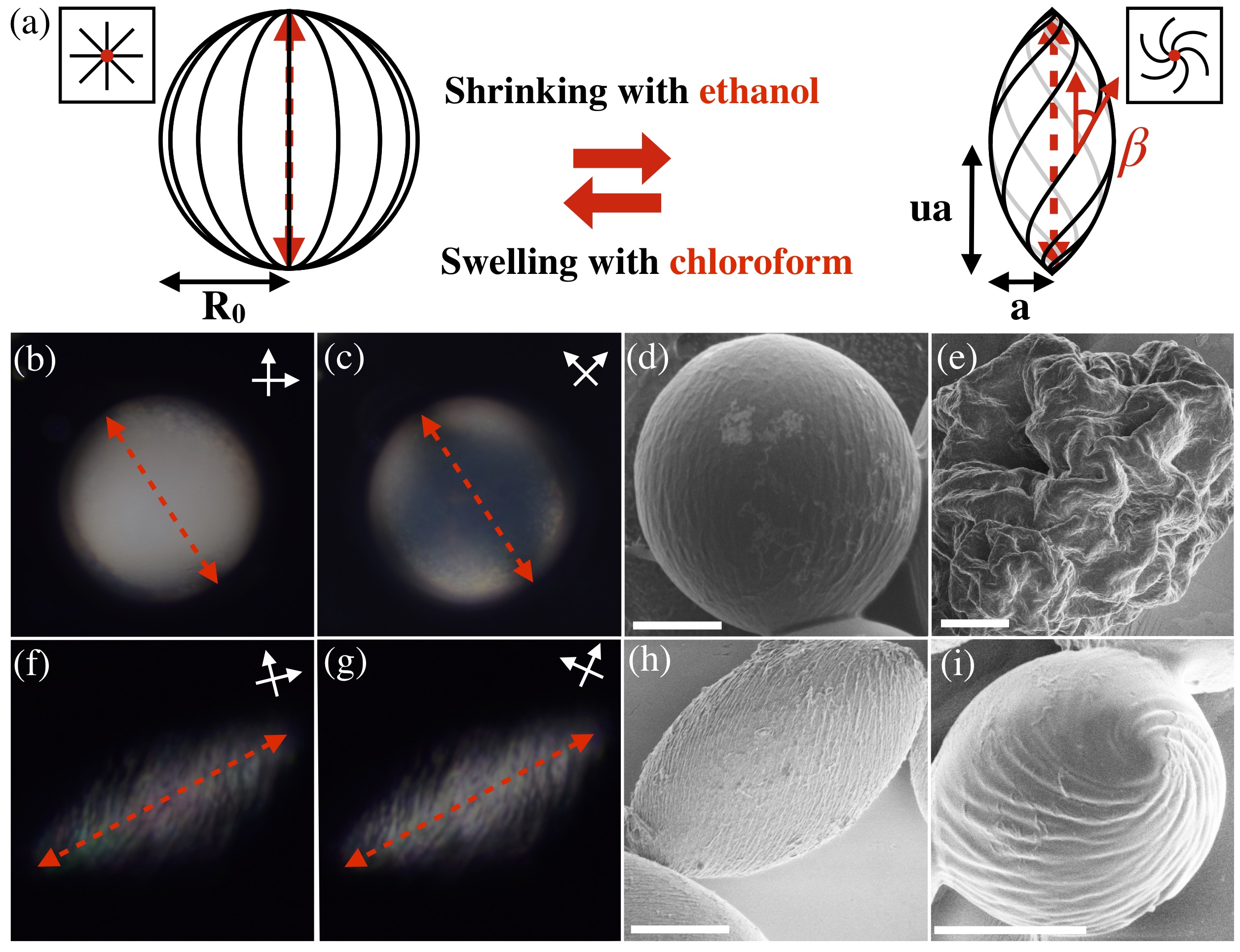}	
		\caption{\label{fig1}(a) Schematic of shrinking and swelling process. The square insets on the top left and top right indicate a top view of the director field at the boojums. (b)-(d) Spherical polymer particle after polymerization and before removal of 5CB. (b)-(c) POM images of particles showing that they are in the bipolar configuration: there is light extinction only when the line connecting the two boojums (optical axis) is aligned with the analyzer/polarizer. (d) SEM image of bipolar configuration. (e) SEM image of crumpling if polymerization takes place in the isotropic phase. (f)-(i) Twisted spindle particle after shrinking. (f)-(g) POM images: there is no complete light extinction at any angle when rotating the sample, indicating that the twist angle changes through the particle. (h) SEM side view. (i) SEM top view. All scale bars are \(20\,\mu m\). Dashed red arrows on POM images indicate the axis connecting the two boojums.}
	\end{figure}

We fabricated polymeric micro-liquid crystal (LC) droplets with diameters of approximately \(100\,\mu\)m using a polymerizable mixture of 4-cyano-4\('\)-pentylbiphenyl (5CB, Merck) and reactive LC monomer 4-(3-acryloyoxy-propyloxy) benzoic acid 2-methyl-1,4-phenylene ester (RM257) dispersed in aqueous solution containing 1wt\% polyvinyl alcohol (PVA) and 2wt\% of photoinitiator (Irgacure 369, Ciba) with respect to the amount of RM257, prepared using microfluidics~\cite{Kim2011}. PVA stabilizes the droplets and enforces planar anchoring at the interface between the liquid crystal and aqueous solution. For most liquid crystals, this type of boundary condition results in the formation of a bipolar configuration, schematically represented in FIG.~\ref{fig1}(a)~(left)~\cite{Drzaic}, where the molecules align in average along the meridians of the sphere, producing two surface defects, or boojums, at the points where the meridians intercept. 

When UV light is shone on the droplets, the RM257 polymerizes in an end-to-end configuration. After polymerization of the RM257, the polymeric particles show preservation of the bipolar configuration, indicating that the majority of RM257 is well oriented along the director field of non-reactive monomers prior to polymerization and that this alignment is preserved during the polymerization process~ \cite{Wang2016,Mondiot2013}. The bipolar configuration is shown in the polarized optical microscopy (POM) images of the particles after polymerization (FIG.~\ref{fig1}(b-c)) as well as in the surface topography shown in the scanning electron microscopy (SEM) image in FIG.~\ref{fig1}(d). Subsequent extraction of the 5CB from the polymeric droplets using ethanol results in pronounced shrinkage of the microparticles in all directions, but with greater shrinking along the directions perpendicular to the line connecting the two boojums. The resulting polymer particles therefore adopt a spindle shape with the defects at the tips, as shown in the schematic in FIG. 1(a)~(right). During the deswelling process, the particles also adopt a twisted pattern as they shrink, as shown in the bottom row of FIG. 1 and in MOV. S1 of the supplementary material. We note that by repeating the same procedure with the droplet in the isotropic phase (approx. \(50^{\mathrm{o}}\)C with 5 wt\% RM257), the resulting polymer network is randomly oriented. Upon removal of the 5CB the remaining polymer particle crumples, as shown in FIG.~\ref{fig1}(e), demonstrating the importance of the initial design of the polymer network in its structure and function. 

We find that when the amount of RM257 is below 20 wt\% we almost always observe twisted spindle structures (see supplementary information). Conversely, if the proportion of RM257 in the initial LC state is too high, the reduction in volume of the resulting particle during shrinking is much smaller and the twisting pattern is not observed, as shown in FIG.~S1 of the supplementary material for a preparation with 40 wt\% RM257. The higher density of the polymer network means that the polymer bundles are closer to each other in the initial configuration. Removing the 5CB creates less free space within the polymer particle into which the polymer can move. Volume reduction through twisting therefore is not preferable in this case. Thus, here, we report and characterize the twisting behavior for 5 wt\% and 20 wt\% of RM257. 

\begin{figure}
		\includegraphics[width=0.48\textwidth]{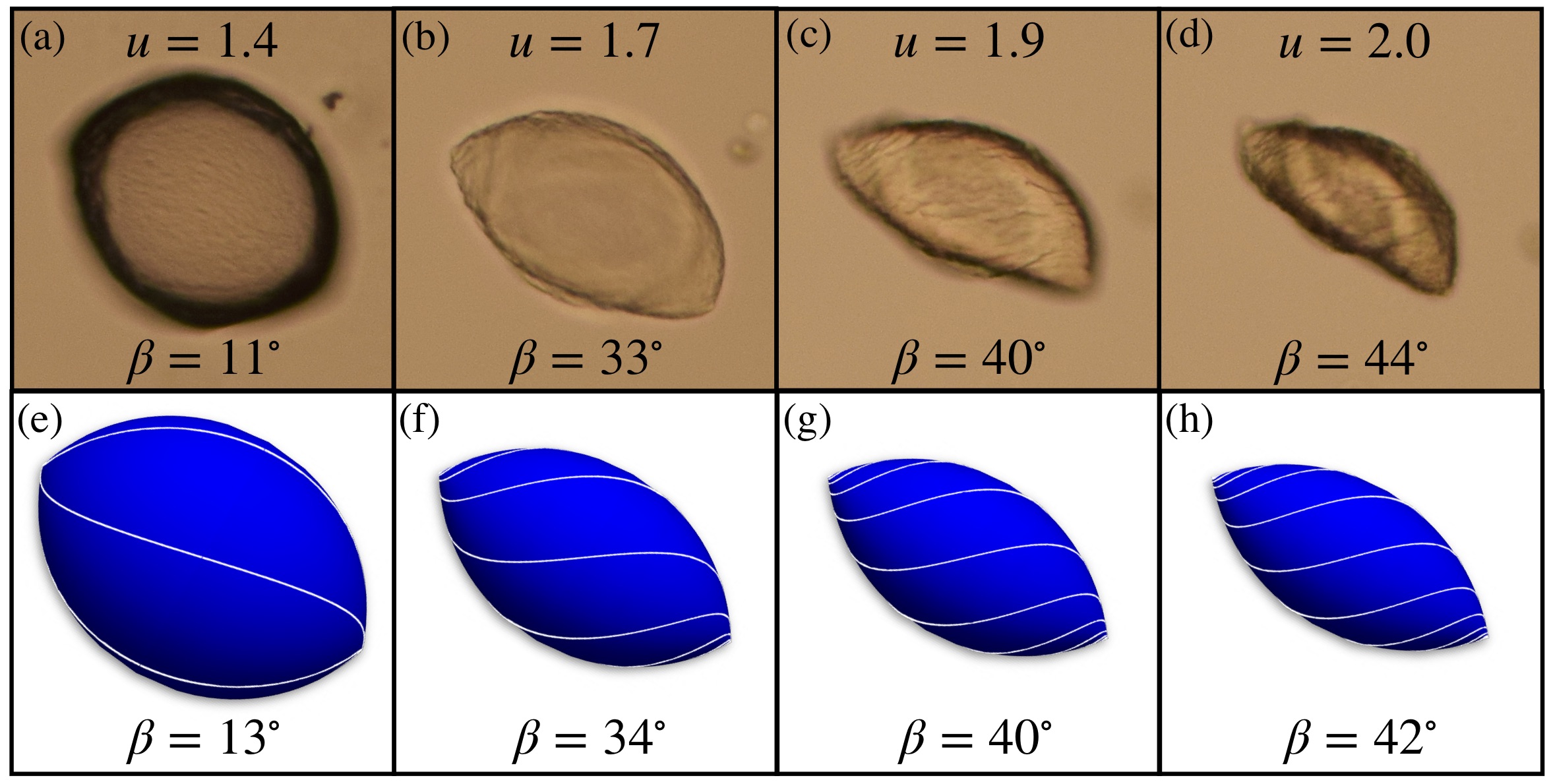}	
		\caption{\label{fig2}(a)-(d) Images of polymer microparticles corresponding to \(\chi_{EthOH} =\) 0, 0.4, 0.6, 0.8 from left to right. The twist angle \(\beta\) and aspect ratio \(u\) increase as \(\chi_{EthOH}\) is increased.  (e)-(h) Corresponding images using the loxodrome model on the spindle for the same aspect ratios. The values of \(\beta\) predicted by the model are given.} 
	\end{figure}

\begin{figure}
		\includegraphics[width=0.48\textwidth]{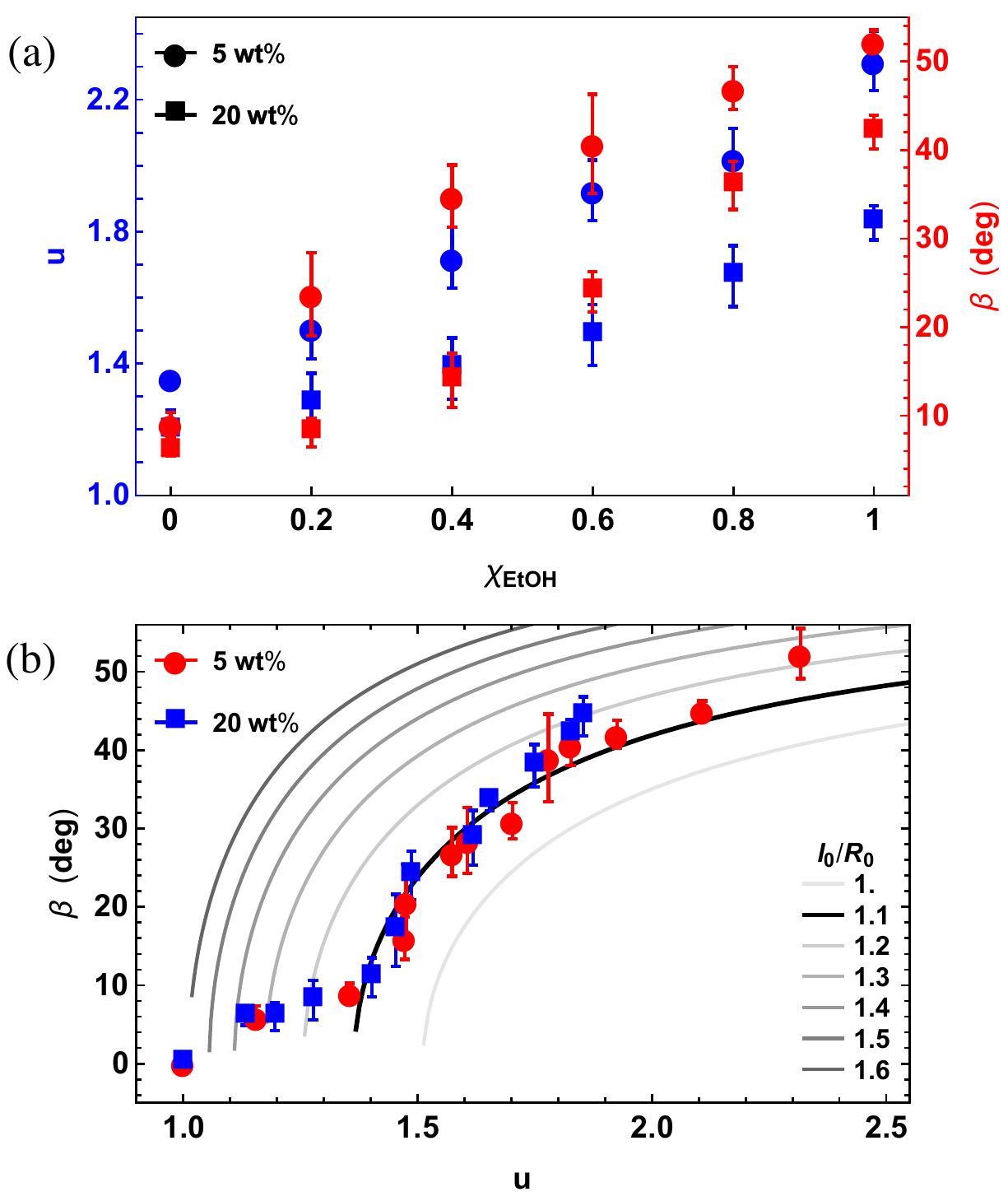}	
		\caption{\label{fig3}(a) Experimental data for how aspect ratio \(u\) and twist angle \(\beta\) vary with  fractional concentration of ethanol. (b) Experimental data and loxodrome model for twist angle as a function of aspect ratio.}
	\end{figure}

We produced equilibrium polymer structures with increasing aspect ratio and twist angle, using a mixture of chloroform and ethanol, in which chloroform and ethanol are swelling and shrinking solvents for poly(RM257), respectively, and studied the relationship between \(\beta\) and \(u\). Experimentally, we measure \(\beta\) as the angle between the meridian connecting the poles of the particle and the lines of the polymer texture at the equator. FIG.~\ref{fig2}(a-d) show bright field OM images of the twisted spindle particles (5 wt\% RM257) in chloroform-ethanol mixtures at equilibrium for different fractional concentrations of ethanol \(\chi_{EthOH}\), ranging from \(u=1.4\) and \(\beta=11^\mathrm{o}\) at \(\chi_{EthOH}=\)0 to \(u=2.0\) and \(\beta=44^\mathrm{o}\) at \(\chi_{EthOH}=\) 0.8. Aspect ratio and twist angle both increase as \(\mathrm{\chi_{EthOH}}\) increases, as shown in FIG.~\ref{fig3}(a), with the growth rate higher for 5 wt\% RM257 than for 20 wt\%.

By plotting the values of \(\beta\) and \(u\) on a single graph, as shown in FIG. \ref{fig3}(b), we observe that the data points for the two different initial polymer densities appear to follow the same behavior and fall on a single curve. The twisting process shows two distinct stages: initially, for \(u<1.35\), there is a small increase in \(\beta\) as \(u\) increases. However, for \(u>1.35\), the rate of change of \(\beta\) with \(u\) increases initially before slowing at larger \(u\). 


In order to understand the observed behavior, we must first consider the structure of the nematic director field in the droplets prior to polymerization. In the bulk, the lines of the nematic director field follow the edges of concentric spindle shapes of increasing minor axis length from the center to the surface of the droplet with their tips aligned at the two defects. After polymerization and removal of the 5CB, the polymer bundles follow the lines of the nematic director field and therefore form a concentric spindle structure, similar to the layered structure of an onion. 

Mathematically, we parameterize the spindles of semi-minor axis length \(a\) and aspect ratio \(u\) using cylindrical coordinates \((r,\theta,z)\), where the major axis of the spindle is parallel to the \(\vu{z}\) direction. The radius of the spindle varies from \(r=0\) at \(z=\pm ua\) to \(r=a\) at \(z=0\). On the surface, the height \(z(r)\) is given by
	\begin{equation}	
		z(r) = \pm a \sqrt{\left(1-r/a\right)\left(u^2+r/a\right)}.
	\end{equation}	
	
There are two distinct ways in which the volume of the particles can change as the solvent concentration is varied. One is for the polymer chains forming the particles to change in effective length due to folding of the chains as the solvent concentration changes. The other is for the polymer chains to rearrange to fit into a smaller volume while maintaining a fixed effective length. We will demonstrate that our results are consistent with both of these mechanisms contributing during the twisted spindle formation.

We now consider the twisting mechanism, for which we consider fixed length polymer chains on the surface layer of the particle. The mechanism presented also applies to each layer within the bulk, resulting in a concentric twisted spindle structure. As the polymer particle shrinks from its initially spherical configuration, each layer must reduce in surface area in order for the overall volume to reduce. Polymer strands, which initially follow meridians, are too long to fit onto the meridians of the slightly smaller spindle surface on the layer below; the strands must rotate in order to accommodate their excess length. One might expect that the polymers would follow geodesics on the surface. However, all geodesics passing through the tip of a spindle are meridians: they do not show any twisting. Since we observe the twisting pattern across the entire surface, we conclude that the curves followed by the polymer strands cannot be only be geodesics.

Although the microparticles shrink anisotropically, the surrounding solvent is isotropic. Consequently, each region of the surface encounters the same local solvent environment. If the twisting pattern is due to each line element of polymer rotating in its tangent plane according to its local environment, and all line elements on a given layer have the same local environment, then all line elements will rotate by the same amount in order to accommodate their length while reducing the overall surface area of the layer. A curve with this property, for which the tangent vector is at the same angle with respect to the local meridian at all points along the curve, is referred to as a \textit{loxodrome}, a term used in navigation to describe a course of constant bearing~\cite{Carlton-Wippern1992}.

A loxodrome of twist angle \(\beta\), defined as the angle between the meridian and the curve (see schematic in FIG.~\ref{fig1}(a)), must have its tangent vector parallel to the unit vector \(\vu{\beta}\) given by
	\begin{equation}					
		\vu{\beta} = \cos{\beta}\,\vu{e}_r + \sin{\beta}\,\vu{e}_{\theta},
	\end{equation}
where \(\vu{e}_r\) and \(\vu{e}_{\theta}\) are unit vectors in the \(r\) and \(\theta\) directions respectively. Satisfying this requirement on the spindle means the curve \(\theta_c(r)\) must obey
	\begin{equation}
		\dv{\theta_c}{r} = \frac{\left(1+u^2\right) \tan{\beta}}{2 r \sqrt{\left(1-r/a\right) \left(u^2+r/a\right)}}.
	\end{equation}	
		
Solving for \(\theta_c(r)\) and choosing the integration constant so that \(\theta_c(a)=0\) leads to the equation of the spindle loxodrome: 
	\begin{equation}
		\theta_c(r) =(u+u^{-1}) \tan{\beta}\tanh ^{-1}\left[\frac{(1-r/a) u^2}{u^2+r/a}\right]^{1/2}.
		\label{eq:theta}
   	\end{equation}

We consider polymer strands along the surface of the particle running from the spindle tip to the equator. If these surface strands have fixed length \(l_0\), the twist angle \(\beta\) as a function of \(a\) and \(u\) can be determined from EQ. \eqref{eq:theta} and is given by  
	\begin{equation}
		\beta = \cos^{-1}\left[\frac{a}{l_0}\left(1+u^2\right)\tan^{-1}\left(u^{-1}\right)	\right].
		\label{eq:beta}
	\end{equation}
Experimentally, we observe a linear relationship between the change in major and minor axis length during the shrinking process (see supplementary FIG. S2), leading to the relation 
	\begin{equation}
		u=0.54R_0/a + 0.48,
		\label{eq:u}
	\end{equation}
where \(R_0\) is the initial radius of the polymer particle (\(\sim50\,\mu m\)). Using this relationship, we can write \(\beta\) (EQ. \eqref{eq:beta}) in terms of the single variable \(u\) for fixed \(l_0\) and \(R_0\). 

During the deswelling the final major axis length of the particles is determined by bulk properties, specifically the decrease in length of a polymer chain directly through the center of the particle connecting the two boojum defects. Experimentally, we observe a 30\% reduction in length of the major axis. Assuming that all strands reduce in length by the same proportion, we can consider a polymer strand along the surface of the sphere and predict the value of \(l_0\). A polymer strand along the surface from the pole to the equator of an initially spherical particle has length \(\pi R_0/2\approx 1.6R_0\). If all polymer strands reduce in length by the same proportion, this strand will have a final length of \(1.1 R_0\) in the twisted state, corresponding to a length of \(\sim55~\mu m\). We therefore use this value as the length \(l_0\) of the polymer strands on the surface.

Plots of \(\beta\) as a function of \(u\), as given by substituting EQ. \eqref{eq:u} into EQ. \eqref{eq:beta}, are shown in FIG.~\ref{fig3}(b) for different values of \(l_0/R_0\). There is good agreement between the model and experimental results for the curve with \(l_0 = 1.1 R_0\) for \(u>1.35\), which is consistent with the loxodrome model with \(l_0 = 1.1 R_0\) being the final length of the polymer strands on the surface. For \(u<1.35\), the behavior of the particles is different. There is not much twisting observed, but rather most of the volume reduction of the particles is due to polymer chains reducing in effective length. The data points lie on curves for larger values of \(l_0\) in the model, which is consistent with the idea of the strands on the surface reducing in effective length during the initial shrinking stages of the particles. 

To understand why this transition from shrinking with minimal twisting to shrinking with more twisting occurs around \(u=1.35\), we consider the known size of a spindle with this aspect ratio using EQ. \eqref{eq:u} and use it to determine \(m\), the length of a meridian from a pole to the equator. For a spindle with \(u=1.35\), the calculated value of \(m\) is \(1.1 R_0\). This value of \(m\) is therefore consistent with \(u=1.35\) being the aspect ratio for which the polymer strands at the surface of the particle have reached their final length. At this value, the twisting mechanism takes over from polymer chain folding as the primary mechanism by which the particles lose volume. This observation, along with the observation that the major and minor axis lengths change continuously during the shrinking process, suggests that the particles shrink from the surface inwards. While strands at the center do not reach their minimum length until \(\chi_{EthOH}\) is maximal, strands near the surface reach their final length earlier in the shrinking process. 


Finally, we would like to highlight that the shrinking and twisting mechanisms described above are fully reversible, providing interesting possibilities for applications. Upon repeated shrinking and swelling cycles in which the particle is swelled in chloroform from the twisted state back to almost a perfect sphere (\(u=1.1\) for RM257 at 20 wt\%) before shrinking back to a twisted state using pure ethanol (\(u=1.8\) for RM257 at 20 wt\%), the aspect ratios achieved in these states are the same after ten repeated cycles. This indicates that the polymer particles have long-term actuating capability and reliability. While it is possible to swell the particles until they are almost perfectly spherical, it is not possible to change the chirality of the twisting through the swelling process. That is, while there is an equal likelihood of a given particle having either handedness, once a particle has twisted in a given direction it will always twist in that direction even after swelling back to almost spherical and shrinking again. This indicates that although the bipolar configuration itself is achiral, there must be fluctuations within the nematic droplets during the preparation process that dictate the chirality of the twisting. 

Since \(\beta\) and \(u\) are highly controllable with solvents, we suggest that these values could be carefully calibrated against the fractional concentration of any binary solvent mixture in which one solvent causes swelling of the polymer particles and the other causes shrinking. The particles could then be used as a quickly responsive and quantitative indicator of the fractional concentration of solvents in the mixture. With further efforts to assemble those twisted particles into three dimensional forms, they may be used as a template in the fabrication of complex three dimensional helical structures for metastable opto-electric devices~\cite{Gansel2009, Yu2014}.

The authors would like to thank K. Dalnoki-Veress and E. Rapha\"el for fruitful discussions.  HSA, DSK, and RDK were supported by NSF MRSEC Grant DMR-1720530 and a Simons Investigator Grant from the Simons Foundation to RDK. TLL and DSK were supported by the French National Research Agency Grant 13-JS08-0006-01. EK acknowledges support by NSF through the University of Pennsylvania Materials Research Science and Engineering Center (MRSEC) (DMR-1720530), the NSF Award PHY-1554887, and the Simons Foundation.

\(^*\) These authors contributed equally to this work.

\bibliography{spindle_sources.bib}

\end{document}